\documentclass[fleqn,usenatbib]{rasti}

\usepackage{pdflscape}
\usepackage{newtxtext,newtxmath}
\usepackage[T1]{fontenc}
\usepackage{graphicx}
\DeclareRobustCommand{\VAN}[3]{#2}
\let\VANthebibliography\thebibliography
\def\thebibliography{\DeclareRobustCommand{\VAN}[3]{##3}\VANthebibliography}

\title[Science from Rubin/Lasair]
{Enabling Science from the Rubin Alert Stream with Lasair}
\author[R. D. Williams et al.]{Roy D. Williams,$^{1}$\thanks{E-mail: roy@roe.ac.uk},
Gareth P. Francis$^{1}$,
Andy Lawrence$^{1}$,
Terence M. Sloan$^{1}$,
\newauthor Stephen J. Smartt$^{2}$,
Ken W. Smith$^{3}$,
David R. Young$^{3}$
\\
$^{1}$Royal Observatory, University of Edinburgh, Blackford Hill, Edinburgh EH9 3HJ, UK\\
$^{2}$Department of Physics, University of Oxford, Keble Road, Oxford, OX1 3RH, UK\\
$^{3}$Astrophysics Research Centre, School of Mathematics and Physics, Queen's University Belfast, BT7 1NN, UK
}
\date{Accepted XXX. Received YYY; in original form ZZZ}
\pubyear{2024}
\begin{document}
\label{firstpage}
\pagerange{\pageref{firstpage}--\pageref{lastpage}}
\maketitle

\begin{abstract}
Lasair is the UK Community Broker for transient alerts from the Legacy Survey of Space and Time (LSST) from the Vera C. Rubin Observatory. We explain the system's capabilities, how users can achieve their scientific goals, and how Lasair is implemented. Lasair offers users a kit of parts that they can use to build filters to concentrate their desired alerts. The kit has novel lightcurve features, sky context, watchlists of special sky objects and regions of the sky, dynamic crossmatching with catalogues of known astronomical sources, and classifications and annotations from other users and partner projects. These resources can be shared with other users, copied, and modified. Lasair offers real-time machine-to-machine notifications of filtered transient alerts.
Even though the Rubin Observatory is not yet complete, Lasair is a mature system: it has been processing and serving data from the similarly formatted stream of the Zwicky Transient Facility (ZTF) alerts.  
\end{abstract}
\begin{keywords}
Software -- Rubin -- Lasair -- transient
\end{keywords}
\section{Introduction}
The Legacy Survey of Space and Time -- LSST; \citep{Ivezic}, using the Charles Simonyi Telescope at the Vera Rubin Observatory, will be a ten-year survey of the changing sky expected to begin operations in early 2025. Of the $\sim$10 million alerts expected each night, most will be the flickering of active and variable objects and wandering of solar system objects. There will also be a few thousand nightly detections of supernovae and active galaxies.

Lasair is designed to filter the massive stream of transients in different ways for different kinds of science, add value to the LSST alerts, and store alerts with their added value. It is available to users as a website\footnote{\url{https://lasair.lsst.ac.uk}} and an API, as well as real-time push notifications that can communicate machine to machine.

The broker is built to process transient alerts rapidly and to make the key decision: {\it is this an object I want to follow up?} LSST alerts will come at a very high rate, and Lasair takes advantage of the design of the distribution system, which issues events in rich alert packets to enable standalone classification. Incoming alerts are judged only on that rich alert packet, without database interaction, leading to a fast and scalable ingestion system. Lasair has been available for some years \citep{SmithWilliams}, as an LSST prototype , delivering alerts from the Zwicky Transient Facility \citep{ztf}.

\section{Lasair Concepts}
The philosophy of Lasair is to be a {\it platform for science} rather than producing the science itself. Lasair doesn't focus exclusively on {\it classifying}; rather, it allows scientists to make their own filters and classifiers from numerical features and attributes. Suppose, for example, a classifier is built to pick out a specific type of supernova. The advantage of Lasair's approach, with user-made features and filters, is that it may be possible to differentiate supernova subtypes, based on a user-defined filter, whereas pre-built classifiers typically lump them together.

Lasair users combine attributes from several rich data tables to build a filter, or they can copy a public filter built by another user and modify it. These rich tables are described below. Filters can be run on-demand on the database of past alerts or in real time, with the results fed rapidly into a machine-readable notification so that a user's own machine can take further action.

Lasair adds to the lightcurve features already computed by Rubin for periodic/stochastic sources, with Lasair features being directed to finding new explosive transients.

Lasair filters can utilise the powerful {\it Sherlock} system \citep{Young_sherlock}: a contextual classifier for astronomical transients that crossmatches a transient's sky-location against an extensive library of astronomical catalogues and, based on matched data, attempts to categorise the transient into one of seven classes; variable star {\tt VS}, cataclysmic variable {\tt CV}, bright star {\tt BS}, nuclear transient {\tt NT}, supernova {\tt SN}, active galactic nucleus {\tt AGN} and orphan (if the transient fails to be matched against any catalogued source). Lasair also has a continuously updated copy of the Transient Name Server (TNS)\footnote{\url{https://www.wis-tns.org/}} database \citep{TNS}; the IAU reporting system for supernovae and other transient sources.

Users can upload their own selection of interesting objects as a {\it watchlist}, which can become a filter of those alerts which are crossmatched with the watchlist. A related resource is the {\it watchmap} to find all alerts that fall into a specific region of the sky. 

Lasair encourages users to build {\it annotator} systems as a way to contribute further value to the information portfolio of an event. Lasair has agreements with other brokers to contribute their classifications as annotations. A user can set up a machine to receive, compute, and contribute information about alerts: an example is NEEDLE \citep{sheng2023neural}, which collects pixels and lightcurves to find rare supernovae and tidal disruption events.

All these resources can be used together in Lasair filters: Lightcurve features, Sherlock, TNS crossmatching, watchlists, watchmaps, and annotations. User-contributed resources can be shared with others, copied and modified, or kept private. Lasair has extensive documentation as text, python notebooks, and how-to videos \footnote{\url{https://lasair.readthedocs.io/}}.

Lasair filters return an initial selection of objects, based on the rich value-added data content, and users can then manually scan the results with the Lasair Marshall Notebook \citep{lasairdox}, or run their own local code on the lightcurve and all the other data about each object. SQL filters can be escalated from static (run on command) to streaming filters, that run whenever new alerts arrive and push the results via email or Kafka. Another approach could use the Lasair API to download results into a user's own marshall system.

\subsection{Scientific goals of Lasair}
Lasair and its partners will provide combined access to the alerts, annual LSST data releases, and external data sources. The joint system will provide a flexible platform that creative users can adapt to their own ends. The science themes are below.

\subsubsection{Extragalactic Transients}
Luminous transients outside our own galaxy include supernovae, kilonovae, tidal disruption events, AGN flare activity, nuclear transients of unknown origin, gamma-ray burst afterglows, stellar mergers, and compact object mergers. All this science requires lightcurves, links to galaxy and redshift catalogues, precise astrometric cross-matching, correlation with high energy information, and multiwavelength cross-matching.

\subsubsection{Multi-Messenger Astronomy}
Lasair responds rapidly to alerts from NASA-GCN \citep{GCN} about gravitational waves, and is able to set up a watchmap (see below) for the sky area. Lasair will collect LSST alerts in that area and allow users to build filters to find the possible optical counterpart. In the future this capability will be extended to gamma-ray bursts, neutrinos, etc.

\subsubsection{Massive Samples of Supernovae}
Lasair will link all transients to a list of likely host galaxies together with their photometric and spectroscopic redshifts. See TiDES, 4MOST, and SoXS in the ``Partners'' section (\ref{partners}) below. 

\subsubsection{AGN, TDE and long-lived transients}
Lasair will allow users to select known AGN, upload their own AGN catalogues, and select flaring events in both active and passive galaxies. This will support the science of tidal disruption events, changing-look quasars, AGN flares, microlensing of background QSOs by foreground galaxies, and unusual long-lived nuclear transients. 

\subsubsection{Stellar Transients}
Most science for variables (typically recurrent and periodic signals) will be achieved with the annual data releases. However, there is a great opportunity in combining alerts with the data releases. Users can discover outbursts or large amplitude variability through the alerts, which link to the data releases and full multi-year lightcurves. Lasair will provide streams of objects matched to known stars and trigger on a particular magnitude variability index.

\subsection{Objects and Sources}
Lasair handles objects and sources delivered by the Rubin Observatory. A source is a detection by the telescope of an object, a collection of pixels on the telescope's light-collection device, which is significantly brighter than it was in the reference imagery taken at the beginning of the survey. A source is detected with a specific narrowband optical filter: LSST uses filters {\it u,g,r,i,z,y} and ZTF uses {\it g,r}.

The brightness of a source in a transient survey is different from that source in a reference sky acquired in the past. Difference flux can be positive or negative, but when expressed as magnitudes, the measurement has two parts: an absolute value converted to magnitudes and a flag to indicate a positive or negative difference. Note that if nothing was detected in the reference sky, then the difference magnitude is the same as the apparent magnitude.

To detect transients, the LSST will trigger on significant (5$\sigma$) sources which are a positive 
excess with a point-spread-function shape measured
on the difference image\footnote{Negative sources with point-spread-function shapes will also be catalogued for variable object selection}. The 5$\sigma$ detection limit is with respect to the noise in the difference image after the reference image has been subtracted. 
When such a detection is made, a data packet called {\tt diaSource} captures the result of the source detection. The Rubin observatory searches for previous transients close to that sky location, and from these, it builds a collective data packet called an {\tt diaObject}. If the transient is associated with a fixed object in the sky, the {\tt diaObject} and its {\tt diaSources} define the lightcurve of that object. The Lasair project is primarily devoted to these fixed sources. For more information on LSST data products, see \citep{lsstdataprod}.

However, if the source detection is from a moving solar system object, there will be no previous detection at that sky position. In a further processing step at the Rubin Observatory, the locations of sources are crossmatched with all known solar system objects. If a match is found, additional data packets are added to the alert, including {\tt ssObject} with properties of the solar-system object and its orbit.

Thus the alerts are organised as a central data packet and a number of associated packets. For astronomical transients fixed in the sky, the central packet is the {\tt diaObject} and the light curve is defined by the {\tt diaSources}, as well as forced photometry and non-detections. For moving objects, there is primarily the {\tt ssObject} and {\tt ssSources}. Lasair is a platform for science involving fixed objects but it also collaborates with the ``Adler'' project, a platform for solar system objects -- more information in the section ``Partners'' below.

\subsection{Partners} \label{partners}
\subsubsection{UK Data Access Centre} \label{dac}
The UK will build and operate an Independent Data Access Centre (IDAC) \citep{UKIDAC}, sized to serve 20\% of the Rubin community. Like the two Rubin-operated DACs, in the US and Chile, this will be a ``Full IDAC'', serving all the Data Release products. Lasair will work hand-in-hand with the UK IDAC, giving longer-term lightcurves and images to back up discovery from the alert stream, facilitating joint data mining projects, and allowing Lasair to serve proprietary data to those of its users who are Rubin Data Rights Holders \citep{rubindatarights}.

\subsubsection{TiDES and 4MOST}
Lasair works closely with the two major ESO projects that will provide tens of thousands of spectra for LSST supernovae. We will coordinate supernova discoveries in Lasair with spectra from the 4MOST multi-fibre spectrometer on the ESO VISTA telescope. We expect to select 35,000 live transients for spectra and obtain spectra of 70,000 host galaxies in the TiDES (Time Domain Extragalactic Survey) \citep{tides}. This will provide the largest cosmological sample of SN Ia, together with a massive statistical sample to understand supernova explosion physics across a range of redshifts and host galaxy masses and metallicities. Lasair will provide both (reproducible) selection and extract the scientific content (type, phase, redshift etc) to re-ingest into the broker for further user exploitation. 

\subsubsection{SoXS}
Lasair also works closely with the UK team responsible for building the science software infrastructure for SoXS \citep{soxs}; a single-shot 0.35-2 $\mu$m spectrometer on ESO’s New Technology Telescope. ESO is fully dedicating the NTT to time domain science, with the schedule being run by the SoXS consortium. We will enable the SoXS marshall to interface with Lasair to select LSST transients for classification and then re-ingest the information and public data for all users to access.

\subsubsection{PESSTO}
Lasair works with the Public European Southern Observatory Spectroscopic Survey of Transient Objects (PESSTO) \citep{pessto}. It is a public spectroscopic survey that began in 2012, classifying transients from publicly available sources and wide-field surveys, and selecting science targets for detailed spectroscopic and photometric follow-up.

\subsubsection{Citizen Science}
Citizen science builds unique and authentic research experiences for the public that directly engage individuals with little or no scientific training or background, thus lowering the barrier for the public to contribute directly to scientific investigations. Lasair already has a strong collaboration with Zooniverse.org, having established a project \citep{Smith} to hunt for rare superluminous supernovae. This will be broadened, and other projects added, in the LSST era.

\subsubsection{Solar System Science with Adler}
The Legacy Survey of Space and Time (LSST) will discover ~6 million new Solar System small bodies, including the comet to which ESA’s Comet Interceptor mission will target \citep{Schwamb_2019}. With each planetesimal receiving hundreds of observations across six filters, LSST will radically transform the view of our planetary system and usher in a revolution for time-domain Solar System science. The Adler system will work closely with Lasair, ingesting the Rubin Solar System alerts and identifying potentially active sources and unexpected photometric behaviour.

\subsubsection{Other Rubin Alert Brokers}
The LSST alert stream is expected to be high bandwidth (0.2 - 5 Gb/sec)\citep{keynumbers}, so it would be impractical to host large numbers of readers, each taking the whole stream. The Rubin Observatory instead selected seven ``Community Brokers'' through a proposal process, each providing its own vision of how such a broker should work. In addition to Lasair, they are
Alerce \citep{alerce},
ANTARES \citep{antares},
AMPEL \citep{ampel},
Babamul \citep{babamul},
FINK \citep{fink},
and Pitt-Google \citep{pittgoogle}.

\subsection{What Lasair is not}
The ``rich data packet'' that comes with each alert means that lightcurve features and subsequent filtering are made only on that packet, which has only a year of past data about each object (only a month for the ZTF prototype). Lightcurve features and thus decisions are {\it not} based on the full multi-year light curve, even though the Lasair databases include these. Rather, decisions are based on these shorter light curves. Even though the lightcurve features are based on the short lightcurve, the web page and API deliver all of it. In the case of ZTF, Lasair lightcurves go back 6 years, and these can be effectively mined -- see section \ref{mining} below.

The fact that a feature is not built into Lasair doesn’t mean it can’t be done; rather, the user has to do some of the work themselves through annotation—and we try to provide a good set of example notebooks to act as starting points for some of these things. It comes down to the distinction between doing science and providing a toolkit for people to do science with.

\section{Lasair Functionality}
Lasair provides a portfolio of data for each object ({\tt diaObject}), and a means to build a filter that selects which objects the user finds interesting. In addition to that provided by the Rubin Observatory pipeline, Lasair adds novel lightcurve features, rich sky context via Sherlock, watchlists and watchmaps, TNS crossmatch, and annotations from other users and brokers. All this can be combined into the {\tt WHERE} clause of an SQL statement that is the heart of a Lasair filter.

The components described below are like LEGO bricks that can be used to build a bespoke filter that concentrates the alerts a scientist is searching for. Each component is expressed as a database table whose attributes can be used to create the filter. Filters can be public and, therefore, shared, copied and modified; the same is true of the other resources described below.

In addition to making filters, users can create watchlists of their special objects and watchmaps of sky areas or build their own classifier running on their own machine. They can then push the results back to Lasair as annotations.

\subsection{Sky Context from Sherlock}
Sherlock finds known astronomical objects at the sky location of an alert. Published astronomical catalogues are carefully curated collections of stars, galaxies, variable stars, cataclysmic variables, active galactic nuclei, etc. When a transient object is detected, an astronomer will want to know if the transient is associated with a previously known source and, if so, what kind of source it is. If astronomers are searching for extra-galactic explosive events such as supernovae, they are typically interested in transients associated with a galaxy. 

Sherlock consists of a Python package and a curated library of astronomical catalogues (in a MariaDB database) and provides a rapid and reliable spatial cross-match service for any astrophysical variable or transient. It associates a transient's position against its library of astronomical catalogues and uses an intelligent ranking algorithm to assign a classification. At its most basic, Sherlock can triage transients into stars, AGN, and supernova-like objects. It has been used for many years as part of the ATLAS, Pan-STARRS, and Lasair-ZTF decision-making.

One of the main purposes of Sherlock within Lasair is to identify known variable stars, since they will make up a majority of LSST alerts, and to associate candidate extragalactic sources with potential host galaxies. Sherlock's library of catalogues contains datasets from many all-sky surveys such as:
\begin{itemize}
\item Gaia DR2 \citep{Gaia}

\item Pan-STARRS1 Science Consortium surveys \citep{panstarrs} and the catalogue of probabilistic classifications of unresolved point sources by \citep{Tachibana} based on the Pan-STARRS1 survey data.

\item The SDSS DR12 \texttt{PhotoObjAll} and SDSS DR12 \texttt{SpecObjAll} Tables \citep{Alam_2015}. The \texttt{PhotoObjAll} table contains a photometry-based measurement of the star-galaxy separation, and the \texttt{SpecObjAll} table contains a spectroscopic redshift for each source.
\item Guide Star Catalogue v2.3 \citep{Lasker_2008}.
\item 2MASS point- and extended-source catalogues \citep{Skrutskie_2006}.
\end{itemize}

Sherlock also employs many smaller source-specific catalogues such as
\begin{itemize}
\item Million Quasars Catalog v5.2 \citep{Flesch_2021},
\item Veron-Cetty AGN Catalogue v13 \citep{VeronCetty},
\item Downes Catalog of CVs \citep{Downes},
\item Ritter Cataclysmic Binaries Catalog v7.21 \citep{Ritter}.
\end{itemize}

For spectroscopic redshifts and/or non-redshift based distance measurements Sherlock uses:
\begin{itemize}
\item LASr-GPS is a 100Mpc volume-limited galaxy catalogue  \citep{10.1093/mnras/staa766} and the
\item NED-D Galaxy Catalogue v17.1.2. A Master List of Redshift-Independent Extragalactic Distances \citep{2017AJ....153...37S}
\item A dynamic query of the NASA/IPAC Extragalactic Database via the \texttt{neddy} software package \citep{Young_neddy}.
\item SDSS DR12 \texttt{SpecObjAll} table \citep{Alam_2015}.
\end{itemize}

At a base level of matching, Sherlock distinguishes between transient objects synonymous with (the same as, or very closely linked to) and those it deems as merely associated with the catalogued source. The resulting classifications are tagged as synonyms and associations, with synonyms providing intrinsically more secure transient nature predictions than associations. Depending on the underpinning characteristics of the source, there are 7 types of predicted-nature classifications that Sherlock will assign to a transient:
\begin{itemize}
\item {\tt VS}: Variable Star, if the transient lies within the synonym radius of a catalogued point-source,
\item {\tt CV}: Cataclysmic Variable, if the transient lies within the synonym radius of a catalogued CV,
\item {\tt BS}: Bright Star, if the transient is not matched against the synonym radius of a star but is associated within the magnitude-dependent association radius,
\item {\tt AGN}: Active Galactic Nucleus, if the transient falls within the synonym radius of catalogued AGN or QSO.
\item {\tt NT}: Nuclear Transient, if the transient falls within the synonym radius of the core of a resolved galaxy,
\item {\tt SN}: Supernova, if the transient is not classified as an NT but is found within the magnitude-, morphology- or distance-dependant association radius of a galaxy, or
\item {\tt ORPHAN}: if the transient fails to be matched against any catalogued source.
\end{itemize}
For Lasair-ZTF the synonym radius is set at 1.5 arcsec, which we find is a good match to the systems PSF and astrometric accuracy. This is the crossmatch-radius used to assign predictions of {\tt VS, CV, AGN} and {\tt NT}. The process of attempting to associate a transient with a catalogued galaxy is relatively nuanced compared with other crossmatches as there are often a variety of data assigned to the galaxy that help to more reliably inform the decision to associate the transient with the galaxy or not. 

Once each transient has a set of independently crossmatched synonyms and associations, Sherlock selects what it thinks is the most likely classification. The last step is to calculate value-added parameters for the transients, such as absolute peak magnitude if a redshift or distance can be assigned from a matched catalogued source. The predicted nature of each transient is presented to the user along with the lightcurve and other information. 

As part of the Lasair project, there is public access to the integrated Sherlock code and database information through the Lasair API \citep{lasairdox}. We plan further additions to the Sherlock database with the Legacy Survey catalogues and eventually LSST data releases.

\subsection{Transient Name Server}
An astronomer who is interested in an explosive event wants to know if it has already been seen and registered by a different survey. The Transient Name Server (TNS)\citep{TNS} is the official IAU mechanism for reporting new astronomical transients, with a focus on extragalactic explosive transients. Once spectroscopically confirmed, new supernova discoveries are officially designated an SN name. Lasair keeps a cache of the database, updated every few hours, which can be used in Lasair filters. 

\subsection{Watchlists and Watchmaps}
Many astronomers study a specific set of existing objects and are interested in transients from those -- for example, active galaxies, galaxy clusters, or star formation regions. A watchlist is a set of named points in the sky, together with a radius in arcseconds -- which can be the same for all sources, or different for each - that is, a set of named cones. It is assumed to be a list of ``interesting'' sources so that any transient that falls within the radius of one of the sources might indicate the activity of that source. Each user of the Lasair system can have one or more watchlists, and can make a filter to be alerted when a transient is coincident with a watchlist source. An ``Active'' watchlist is one that is run every day so that it is up to date with the latest objects.

To be specific, suppose we are interested in the 42 objects in the catalogue of BL Lac candidates for TeV observations \citep{massaro}, which can be found in the Vizier library of catalogues. The user can make their watchlist ``public'' so other Lasair users can see it and use it in queries, and can make the watchlist ``active'', meaning that the crossmatch is kept up to date with incoming alerts, and active filters respond immediately. Full details are in \citep{lasairdox}.

A watchmap is a specification of an area of the sky, that can be used as part of a Lasair filter. The crucial purpose here is for gravitational wave events and other multimessenger transients, where the sky location is actually a probability distribution. However, the MOC doesn't embody the idea of a probability distribution -- it's just an area of the sky -- so the watchmap might cover the 90-percentile. A watchmap might also be the footprint of another survey or a large sky area like the Orion Nebula. If the watchmap has been set to ``active'', then all alerts ingested to Lasair are tested against it, and tagged if inside the watchmap. A filter can then be built that selects only alerts inside the watchmap. Watchmaps are created by building and uploading a MOC (Multi-Order Coverage) file \citep{moc}.

\subsection{Annotations}
Lasair allows users to add information to the database, that can then be used as part of a query by another user. An annotation is a structured packet of extra information about a given object, that is stored in the annotations table in the SQL database. This could be the result of running a machine-learning algorithm on the lightcurve, the classification created by another broker, or data from a follow-up observation on the object, for example, a link to a spectrum. Users that put annotations into the Lasair database are validated, and administrators then make it possible. That user will run a method in the Lasair API -- from their own machine -- that pushes the annotation: all this can be automated, meaning the annotation may arrive within minutes of the observation that triggers it.

Each annotation is associated with a specific Lasair object, and may contain:
\begin{itemize}
\item {\tt objectId}: the Lasair object being annotated
\item {\tt topic}: the name of the annotator that produced this annotation
\item {\tt classification}: a short string drawn from a fixed vocabulary, eg “kilonova”.
\item {\tt explanation}: a natural language explanation of the classification, eg “probable kilonova but could also be supernova”
\item {\tt classjson}: the annotation information expressed as a JSON dictionary
\item {\tt url}: a URL where more information can be obtained, for example a spectrum of the object obtained by follow-up.
\end{itemize}

The NEEDLE project \citep{sheng2023neural} is an example of an annotator, designed to find superluminous supernovae in dwarf galaxies, and tidal disruption events occurring in the centres of nucleated galaxies. There is a ``pre-filter'' that selects based on the Sherlock classification and the current magnitude versus mean magnitude; results of this filter are pushed via Kafka to a separate analysis machine, which pulls the lightcurve and cutout images for analysis by a neural-net system. Promising candidates are pushed back to Lasair as an annotation, with the {\tt classjson} attribute set to something like:

\begin{verbatim}
{"SN": "0.799", "SLSN-I": "0.103", "TDE": "0.099"}
\end{verbatim}

These returned attributes can then be used as part of a filter in the usual way; there is special syntax to refer to the parts of the {\tt classjson} in the SQL language.

Another annotator is {\it Fastfinder} that models lightcurves of new transients to find rapid brightening from kilonovae. It classifies alerts as {\tt SLOW}, {\tt FAST}, and {\tt SN} so that the latter could be picked out with a clause {\tt fastfinder.classification="SN"} in the {\tt WHERE} part of the filter. As explained above, the {\tt classjson} can hold complex information, and querying is more sophisticated; for more information see the Lasair documentation \citep{lasairdox}.
Annotations can be pushed to the Lasair database using the Lasair client, however the user must be authenticated to do so. Lasair staff are happy to receive a request to create an annotator, and the successful user's topic name and API key will allow them to upload annotations.

Because annotation is a process of inserting data into the Lasair database, the Lasair Team needs to know who is doing it. Therefore, running an annotator starts with asking the team to set it up, and upload of annotations can only be done by the person responsible. 

\subsubsection{Fast Annotations}
Some special annotations can be upgraded to ``fast'' annotations. This means that as soon as the annotation is uploaded, all user filters that involve that annotator are rerun immediately. Suppose, for example, a user builds a filter based on the NEEDLE annotator, the criterion being that the {\tt SLSN-I} probability for that object is more than 0.7. If NEEDLE is a fast annotator, then that user's filter will run as soon as the annotation arrives, with an immediate email or Kafka result. Otherwise, there will be a delay, until another detection of that object occurs, which may be days later.

\subsubsection{Alerce and Fink} \label{alerce_fink}
Lasair is one of seven Community Brokers for the Rubin alerts and has arranged with Alerce \citep{alerce} and Fink \citep{fink} to consume their outgoing Kafka streams and ingest the results of their classifiers as annotations. Currently, these include the Alerce Stamp Classifier and Light Curve Classifier, as well as some of the many Fink classifiers, which include Kilonova, TDE, Microlensing, and subtypes of supernovae.

\subsection{Mining the Lasair Database -- an example} \label{mining}
Here we provide a science example of how the Lasair database can be mined. There is a `pre-filter', then all the resulting lightcurves are fetched by API and cuts applied, until there are few enough to look by eye.
Our example is based on the transient known as AT2021lwx, which released $1.5 \times 10^{53}$ erg of energy over three years, more than any other optical transient. It was a transient from the ZTF survey but sat in the Lasair database until it was found \citep{wiseman}. AT2021lwx was extragalactic, and had a long, smooth decline in brightness over three years. Below we sketch a way to find other such events in the Lasair database.

For this search, there were two Lasair filters: one for transients in a catalogued galaxy, and another for those where the host galaxy is too faint to have been catalogued. The first query finds alerts with Sherlock classifications {\tt SN}, {\tt NT}, or {\tt AGN}; the second finds Sherlock {\tt ORPHAN}s away from the Milky Way [abs(galactic latitude) < 10]. Each of the resulting 59,000 ZTF lightcurves is fetched from Lasair and features computed for both {\it g} and {\it r} filters: being a linear fit to the lightcurve after the peak brightness. If the variance is low about this line, and the rate of fading is slow, then we have a candidate. The result of this specific case produced 58 candidates, still under investigation (Wiseman, in prep.)

A data-mining project of this kind requires a lot of API calls, therefore requesting that the Lasair team remove the default API throttling -- which is only 20 API calls allowed in any given hour.

\subsection{Realtime Output}
Alerts can be visualised on the Lasair web interface, or the underlying data fetched using the Lasair API. But there is another way to get Lasair data, through {\it push} notification, where the Lasair system pushes data as soon as it is available. The push can be a human-readable email, or via a machine-readable Kafka stream. Such a stream is designed to be consumed by machines, not people; it may be that an analysis program sits waiting, and springs into action as soon as the Kafka alert arrives. It could compute a lightcurve classification, it could drive a followup telescope, perhaps followed by an annotation. The consumer of a Kafka stream need not run continuously as alerts can build up until they are summoned, perhaps by the Lasair Marshall Notebook (see section \ref{marshall}). For more information about Lasair's Kafka output, see section \ref{realtime}.

\section{Lasair in Practice}
\subsection{Making a Filter}
\begin{figure}
\includegraphics[width=8cm]{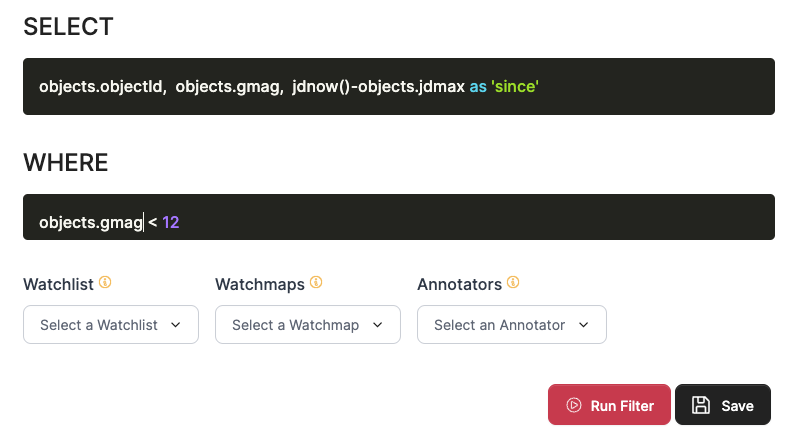}
\caption{Screenshot of the filter builder showing the user inputting the {\tt SELECT} and {\tt WHERE} parts of the filter.}
\label{fig:brightstars}
\end{figure}

Lasair is built around ``filters'' of the alert stream. Users create a filter with SQL clauses, based on the attributes of the object associated with the alert; its lightcurve, sky context, etc. First, the user makes a filter and runs it on previous alerts, and then they can save the filter. The user can convert a filter to an ``active'' filter, so that results are sent to email or to their own machine as soon as they are available.

In Figure \ref{fig:brightstars} is a screenshot of an example filter for bright stars that have had recent alerts. The {\tt SELECT} clause can be typed (with the assistance of auto-complete and schema browser) as:
\begin{verbatim}
objects.objectId, objects.gmag, 
jdnow()-objects.jdmax AS since
\end{verbatim}
and the {\tt WHERE} clause similarly as:
\begin{verbatim}
objects.gmag < 19 ORDER BY objects.gmag 
\end{verbatim}
In English, the filter is: select the object ID, g-magnitude, and time in days -- call it {\tt since} -- since the latest detection, and we only want to see those objects brighter than 19th magnitude, sorted with the brightest first. There is also a {\tt FROM} section listing the tables being joined to make the query; this list is made automatically from the content of the {\tt SELECT} and {\tt WHERE} clauses.

Clicking ``Run Filter'' gives the brightest from years of stored alerts, as expected. If, however, the filter is saved, there is a popup form asking for name, description, etc, as well as if it should be public and if it should be ``active''. The active filter is one of Lasair's crucial and powerful ideas: it means that all incoming alerts are passed through the filter automatically, in near-real-time, and what passes the filter is pushed to the user, by email or by Kafka, which can also drive a Slack workspace or other alert mechanism. Filters run the same way whether the filter operation is on-demand by click/API, or whether it runs in near-real-time. However there is one difference: the above filter run on-demand returns the brightest alerts first (because of the {\tt ORDER BY} clause), but by the nature of a streaming filter, it can only return all its results in time order.

\begin{figure} \label{watchlist_filter}
\includegraphics[width=8cm]{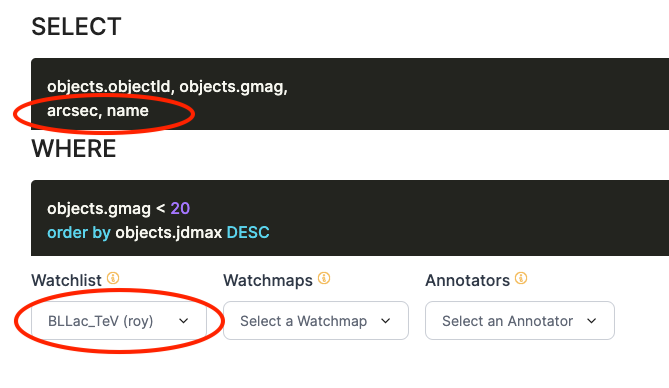}
\caption{Screenshot of the filter builder showing the user inputting the {\tt SELECT} and {\tt WHERE} parts of the filter.}
\label{fig:watchlist_filter}
\end{figure}

To get alerts coincident with a watchlist, Figure \ref{fig:watchlist_filter} shows how all that is needed is choosing the watchlist in the pull-down list. The attributes {\tt arcsec} and {\tt name} are then available for the name of the coincident object from the watchlist, and its distance in arcsec from the alert.

There will be many lightcurve features from the LSST data pipeline, packaged with each alert \citep{Bellm}. Some ideas for adding value to these, now in development, are described in Appendix A.

\subsection{Learning}
Lasair has full documentation \citep{lasairdox}, explaining the basic concepts of Lasair, as well as detailed how-to tutorials. There are also videos showing how to get a user account, how to make watchlist and watchmap, and other topics. On the Lasair webpage itself are numerous tooltips -- short popup explanations when the mouse hovers -- each of which has a link to more detail in the main documentation. There are Jupyter notebooks in the documentation, showing simple and complex queries through the API, the API interface to Sherlock, how to utilise Kafka streams and other topics.
\begin{figure*}
\includegraphics[width=14cm]{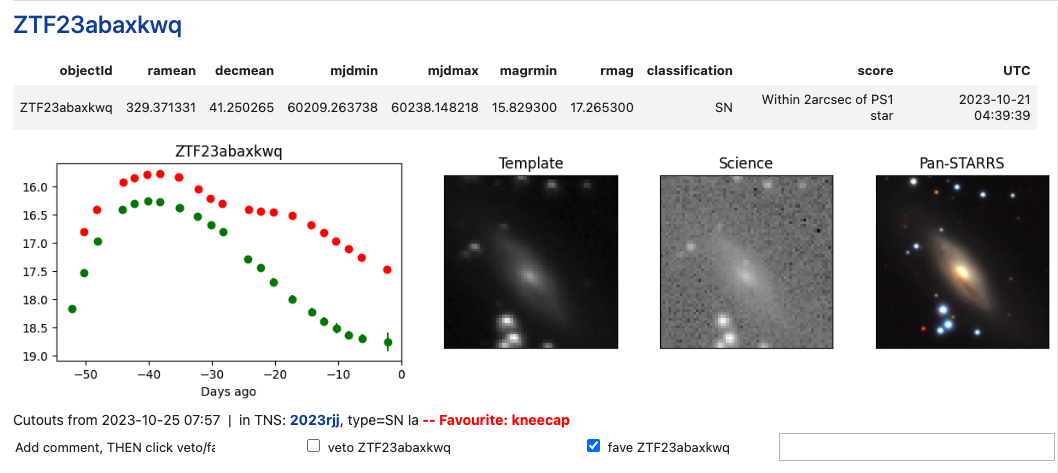}
\caption{Screenshot from the Lasair Marshall Notebook}
\label{fig:marshall}
\end{figure*}
\subsection{The Lasair Client} \label{API}
There is a Python software package called {\tt lasair} \citep{lasair_client} which can be installed with {\tt pip}. It can be used to query the database and Sherlock, to create an annotation, or to consume a Kafka stream. Full details are in the Lasair documentation \citep{lasairdox}.

When a filter has been set to ``active'' and Kafka output, a topic name will be displayed on the web page of that specific filter. It is this that is used to select the output stream from that filter. Users need to understand the concepts of {\tt topic} and {\tt groupId}, perhaps by watching the relevant videos in the documentation. The code then creates a {\tt consumer} object and calls the {\tt poll} method to get the messages. Full details are in the Lasair documentation \citep{lasairdox}.

\subsection{Real-time Notification} \label{realtime}
As described above, the Lasair broker sends notifications when an active filter sees an interesting alert. This can be done by email, which is suitable for low-volume filters and human attention. Notifications can also be sent via Kafka, an open-source distributed event streaming platform used throughout industry for high-performance data pipelines, streaming analytics, data integration, and mission-critical applications (see kafka.apache.org). Indeed, the data ingestion from Rubin and the Lasair processing pipeline are both based on Kafka. Other methods of notification are then easily implemented (e.g. Slack).

\subsection{Marshall Notebook} \label{marshall}
The purpose of a transient survey is to find the needle in the haystack: the scientifically interesting alert among millions of others. Lasair users can do this in two parts: first an automated filter that runs as the alerts are ingested, then a human looking at the results, with an application known as a ``marshall''. The scientist looks at a batch, flagging some as ``favourites'', and some as ``don't show me this object again'' (veto), and checking full objects information for some. Then the next batch, until all have been seen. The next night there will be a fresh batch to be checked.
The Lasair project has built a ``marshall'', implemented as a Jupyter notebook. The alert in Figure \ref{fig:marshall} is a supernova in the centre of the Science image – near the centre of the spiral  NGC 1086.

At the top are the attributes selected by the filter, then lightcurves, the ZTF reference and latest image, and the colour image from Pan-STARRS. The SN II is already registered in TNS, and has a link. There are two checkboxes and a place for a comment. Ticking ``veto'' means the object will not be seen again (like the object at the bottom of the image), and ticking ``fave'' means it will be emphasised next time. If the notebook is run again, a new set of results is shown, until all have been seen. Using the notebook for eyeballing is much easier than having Lasair send email notifications. Further instructions can be found in the Lasair documentation \citep{lasairdox}. This Jupyter notebook is meant for individual users, but we expect that other projects (e.g. SoXS) will build their own marshall system, as the PTF, ZTF, and PESSTO projects have done, connecting through Kafka and the Lasair API.

\subsection{Kinetic Display}
\begin{figure}
\includegraphics[width=8cm] {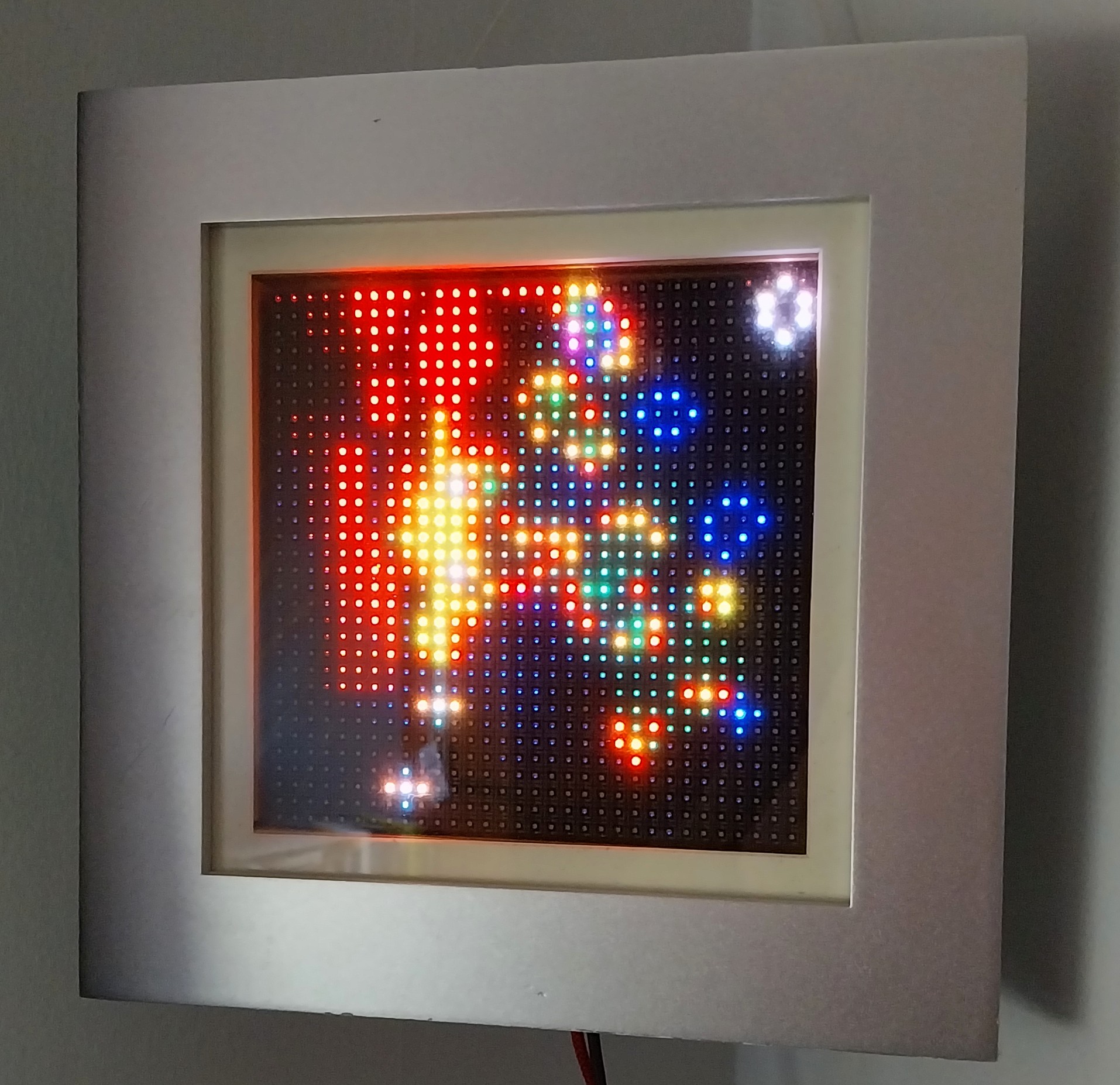}
\caption{The do-it-yourself realtime data display for Lasair's public Kafka}
\label{fig:RGBmatrix}
\end{figure}
This display shows the status of the Lasair data flow, as shown in Figure \ref{fig:RGBmatrix}. The total cost is about £100. A 32x32 RGB matrix is powered by a Raspberry Pi, and it can be mounted in a box frame to make a wall-mounted display. The RPi needs to be on a Wi-Fi network. The code for the display and full instructions can be found in \citep{RGBmatrix}.

Each alert that arrives has RA and Dec in the sky: RA is 0 to 360 from left to right, and Dec is -25 to 90 from bottom to top. Underlying the colours of the display is a matrix of bits. When a batch of alerts has come in, the matrix is replaced with the new matrix with bits switched on based on the positions of the alerts of the batch. Simultaneously, Conway's Game of Life \citep{Conway} cellular automata runs continuously on the matrix, leading to patterns that evolve and change rapidly after the batch of alerts is received. The actual colour display is an RGB matrix derived from the history of the binary matrix, with R, G, and B fading at different rates.

The display can also be shown on a computer monitor, without buying extra hardware, by installing the {\tt OPenCV} \citep{OpenCV} Python package.

\section{Technical Implementation}
\begin{figure}
\includegraphics[width=8cm]{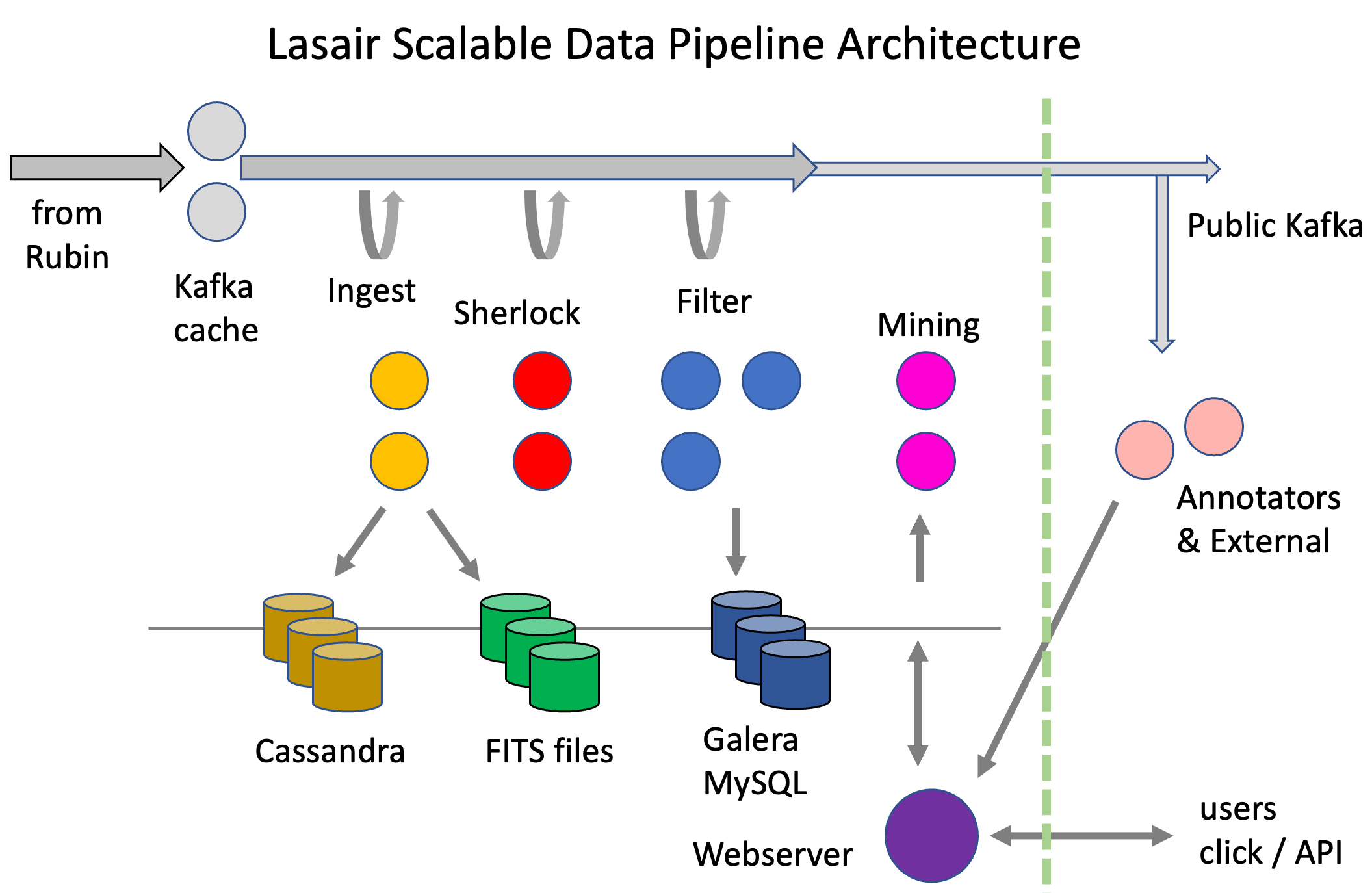}
\caption{The architecture of the Lasair system. Scalable homogeneous clusters are connected by Kafka to form a pipeline}
\label{fig:arch}
\end{figure}

In this section, we give a brief overview of the technical implementation of the Lasair system. The overall architecture is illustrated in Fig. \ref{fig:arch}. Given that Rubin will produce ten million alerts per night, the processing system must be carefully architected and well-resourced to keep up. We don't want Lasair falling hours behind, still processing a night's alerts well into the next day, or even worse, not being finished when the firehose starts up the next night.

Lasair runs on an Openstack cloud ``Somerville'' at the University of Edinburgh Advanced Computing Facility, which is part of ``IRIS'' -- the UK academic cloud \footnote{IRIS: Digital Research Infrastructure for UK Science, \url{https://www.iris.ac.uk/}}. It shares that cloud with the Data Access Centre described in section \ref{dac}.

Lasair ingests data with a pipeline of scalable clusters, as shown in Figure \ref{fig:arch}: Kafka, ingest, Sherlock, filter. Each cluster does a different job, some more compute/data intensive than others. It is difficult to know {\it a priori} how much resource should be allocated to each cluster, so our design gives flexibility: each cluster can be grown or reduced according to need. Also, there are persistent data stores (Cassandra, MariaDB via a Galera cluster); again, each is a resilient cluster architecture that can be grown or reduced according to need. The diagram shows the concept: data enters the Kafka system on the left and progresses to the right. The Kafka cluster (grey) consumes and caches data from Rubin and from the other pipeline clusters; the ingest cluster (yellow) splits and redirects, then puts data back into the Kafka communication bus; the Sherlock cluster (red) consumes and produces Kafka; the filter cluster (blue) consumes the result and writes them to the Galera database cluster and public Kafka. We also include the web and annotator nodes in this picture (bottom and right), as well as the mining nodes, although they are not part of the data ingestion pipeline. 

The web server supports users by delivering web pages and responding to API requests. The annotator nodes may be far from the Lasair computing centre and controlled by Lasair users, but they are in this picture because just like the others, they push data into the data storage and may consume from public Kafka.
The Kafka system is represented by the green nodes in Figure \ref{fig:arch} as well as the grey arrow at the top. It is responsible for reading and caching the alert packets from Rubin, as well as sending data to the compute clusters and receiving their resulting packets.

The Ingest nodes read the original alerts from the Kafka system and put the cutout images in a dedicated database. The recent lightcurve is then sent to the NoSQL (Cassandra) database. The alert without cutouts is reformatted as JSON since there is no binary content, and then it is pushed into the Kafka system.

Each Sherlock node has its own database of 5 Tbytes of astronomical catalogues. If necessary, this can be replicated to provide a higher throughput.

Each filter node computes features of the 12-month LSST light curve that comes with the alert, as well as matching the alert against user-made watchlists and watchmaps. Records are written to a local SQL database onboard the node for the object and features, the Sherlock data, the watchlist and watchmaps tags. Other tables have already been copied into the local database from the main SQL database. After a batch of perhaps 40,000 alerts is ingested into the local database, it can now execute the user-made queries and push out results via the public Kafka system -- or via email if the user has chosen this option. 

The tables in the local database are then pushed to the main SQL database and replace any earlier information where an object is already known. Once a batch is finished, the local database tables are truncated and a new batch started.

There are also ``twilight services'' to prepare the ingestion system for the night: caching annotations, watchlists, and watchmaps for the filter nodes, the latest version of the TNS database, and others.

\subsection{Scalability and Efficiency}
The Lasair project will store all the alert information from Rubin in a Cassandra database (although the more recent cutouts may only be stored there).
Each Rubin alert is issued because of a single detection that is $5\sigma$ significance above the noise in the
difference image (target image minus the reference sky), but it is accompanied by a host of additional data to allow judgement to be made. In principle, this extra data is not needed since it is already available in the Cassandra database; however, this redundancy allows the alert processing to be handled by many independent nodes, with no need to communicate with each other or read from the databases. Thus during data ingestion, there is no reading from the databases at all. When writing data, these systems can work asynchronously without blocking, an ``eventual consistency'' paradigm that allows maximum write speed. Any processes that require consultation with the database are then left to the next day, when the only usage of the databases is much lighter: only the web/API customers.

There is a multi-node relational database behind the web server and API, providing users with the richness of the SQL language. The much large Cassandra database is indexed on {\tt diaObjectId} and sky position, so it is very much a key-value retrieval system rather than a query-based relational database.

\subsection {Webserver}
The Lasair webserver is built in Python with Django to separate computation from presentation. The Twitter Bootstrap v5.2 framework is used alongside Plotly for lightcurve display. AladinLite \citep{aladinlite} shows sky context in numerous wavebands, and the JS9 package is used to display image stamps. A Grafana dashboard shows the status of all the nodes, with live traces of the pipeline's status in terms of alerts per minute.

\subsection{Self-protection}
The Lasair Team has worked to ensure flexible and useful filtering of alerts for users and has chosen the SQL language for this, as it is well-known now to astronomers who are data specialists. However, Lasair guards against mistakes and attacks through SQL. All queries are run in a user account with read-only access, and query strings are carefully vetted before being run against the database. User-made queries have a time-out of 10 seconds when running in streaming mode against a batch of alerts, so they can’t hold up the entire system if they are too resource-intensive. 

The filter nodes compare each alert against active watchlists and watchmaps and run each batch against active user queries. Each of these active resources slows the ingestion system. Therefore, Lasair has a 6-month expiry for each, and the user must then go to the website and renew its activity.

\subsection{Deployment}
Lasair has a flexible system for deploying an entire system on the Openstack cloud, using Ansible and Terraform, with a specification of the numbers of nodes to be used for each functional cluster, their memory and local disk, spinning disk or SSD, how they are to be located on hypervisors, and other hardware considerations. This makes it easy to have extra systems running the entire Lasair stack: for example a development system or a system for measuring performance.

\section{Summary and Conclusion}
Lasair has been operating for years with the ZTF alert stream, has several hundred user accounts, and is ready for the greater challenge of LSST, with about 30 times the alert rate. A user of Lasair does not simply take a ready-made sub-stream of classified alerts; rather, they combine different elements from a kit. The kit has the Sherlock association engine to find what is already published about a specific sky location; it has watchlists, so a user can ask for alerts associated with a set of astrophysical sources; it has watchmaps to restrict attention to arbitrary sky regions; it has sophisticated features to characterise the lightcurve. Lasair allows users to do deep analysis of objects on their own machine and annotate Lasair with their computed results and classifications. These different resources are then combined, using the SQL language, to create a filter, and the filter can run from a web-click, by API, or run in real-time streaming mode as alerts come in, so no time is lost, with machine-readable output. A user can publically share (or not) their watchlists, watchmaps, annotations, and filters. Lasair is being upgraded to tackle gravitational-wave events, thus allowing users to pick the LSST alert most likely to be associated with the counterpart.

\section*{Acknowledgements}
Lasair is currently supported by the UKRI Science and Technology Facilities Council and is a collaboration between the University of Edinburgh, Queen’s University Belfast, and the University of Oxford (grants ST/X001334/1 and ST/X001253/1) within the LSST:UK Science Consortium. Lasair is hosted by the STFC IRIS academic cloud.
Lasair relies on the ZTF survey, supported by the National Science Foundation under Grant No. AST-2034437 and a collaboration including Caltech, IPAC, the Weizmann Institute for Science, the Oskar Klein Center at Stockholm University, the University of Maryland, Deutsches Elektronen-Synchrotron and Humboldt University, the TANGO Consortium of Taiwan, the University of Wisconsin at Milwaukee, Trinity College Dublin, Lawrence Livermore National Laboratories, and IN2P3, France. Operations are conducted by COO, IPAC, and UW. The ZTF forced-photometry service was funded under the Heising-Simons Foundation grant 12540303.
This research has made use of the NASA/IPAC Extragalactic Database (NED), which is funded by the National Aeronautics and Space Administration and operated by the California Institute of Technology.

\section*{Code and Data Availability}
The Lasair software is available at \citep{lasairgithub}.
ZTF data are available at \citep{ZTFdata}. LSST data rights are described at \citep{rubindatarights}.

\bibliographystyle{rasti}
\bibliography{lasair}


\appendix

\section{Lightcurve Features}
In addition to other object properties, lightcurve (photometric) features are computed on each object, and users can build filters based on these. The Rubin observatory will precompute many features \citep{Bellm}: a set of features that comprehensively covers long-lived periodic and stochastic lightcurves. But some of the added value of Lasair is extra lightcurve features tailored to explosive events -- emphasising very recent behaviour -- either a long-term lightcurve that brightens significantly, or a new transient that has appeared recently. 

Non-parametric features have no underlying model; rather, they are statistics, such as mean, median, max flux, moving averages, rate of change of flux, and so on. Lasair includes these and also has a feature {\tt fluxJump} which crudely looks for brightening: first take all fluxes together and compute the standard deviation of this early flux, then compute how many standard deviations different is the current flux. If a user wants to know about sudden brightening, this a first criterion.
\begin{figure} 
\includegraphics[width=8cm]{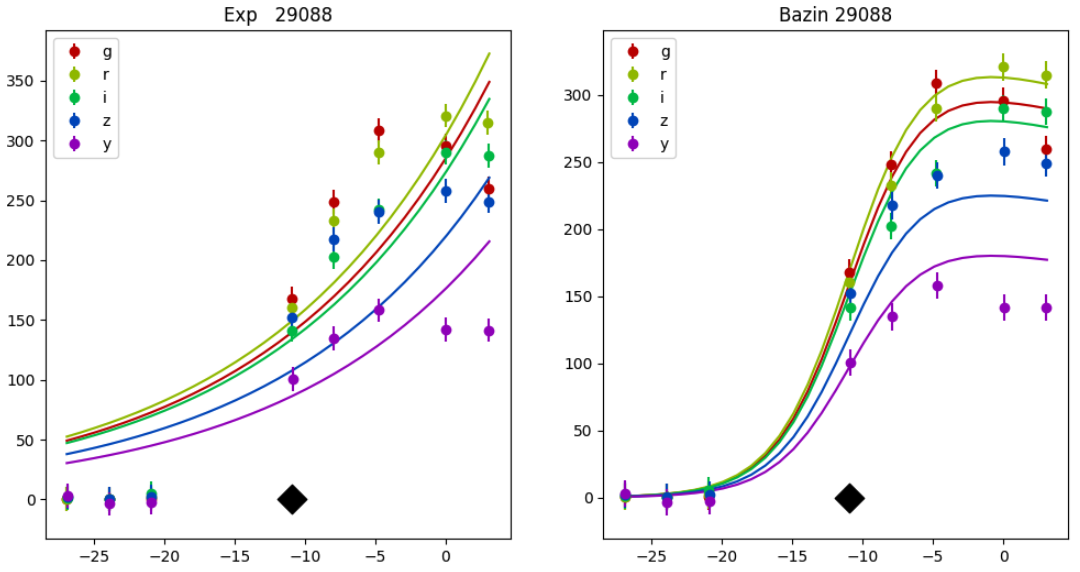}
\caption{Early light curve simulated in six filters, fitted with both exponential (left) and
Bazin (right) for the time dimension and a constant-temperature black-body in the wavelength dimension. In this case, the best-fit criterion chooses the Bazin fit.}
\label{fig:bazinexpbb}
\end{figure}
Parametric features imply fitting a model to the data. In the time dimension such fits could include generic models like linear, polynomial, or exponential, as well as more physically motivated models such as microlensing, eclipsing binary, or Bazin fits \citep{bazin}. Many proposed algorithms work with a monochromatic lightcurve, or use the word ``colour'' to imply there are only two filters whose fluxes can be subtracted. But LSST provides six filters, so perhaps we should not simply build separate features for each filter, but think of the ``time-wavelength surface''. Lasair utilises this 2D space to build a simultaneous black-body fit with either a Bazin curve or a plain exponential in time, as in \citep{bazinexpbb}. Figure \ref{fig:bazinexpbb} shows two fits to some simulated six-filter fluxes, on the left is a product of exponential rise in time and black-body in wavelength, on the right is Bazin in time with black-body. The diamond marks the discovery epoch. A goodness-of-fit criterion picks the best fit as the right-hand plot. This two-dimensional approach to fitting has been used later in other work \citep{rainbow}.


\begingroup
\renewcommand{\section}[2]{}
\endgroup
\label{lastpage}
\end{document}